\documentclass[aps,twocolumn,superscriptaddress,floatfix,showpacs,amsmath,longbibliography,nofootinbib,amssymb]{revtex4-1}
\usepackage[colorlinks=true,citecolor=blue,linkcolor=blue]{hyperref}
\usepackage[usenames]{color}
\usepackage{subfigure}
\usepackage{dcolumn}
\usepackage{mathrsfs}
\usepackage{bm}
\usepackage{amsmath,amssymb,epsfig,float,graphics}
\newcommand{\bk}{\mathbf{k}}
\newcommand{\br}{\mathbf{r}}
\newcommand{\bR}{\mathbf{R}}
\newcommand{\cA}{\mathcal{A}}
\newcommand{\cH}{\mathcal{H}}
\DeclareMathOperator{\erf}{erf}
\DeclareMathOperator{\arctanh}{arctanh}

\begin{document}
\baselineskip=0.45 cm

\title{Probing low-energy Lorentz violation from high-energy modified dispersion in dipolar Bose-Einstein condensates}

\author{Zehua Tian}
\email{tianzh@ustc.edu.cn}
\affiliation{Hefei National Laboratory for Physical Sciences at the Microscale and Department of Modern Physics, University of Science and Technology of China, Hefei 230026, China}
\affiliation{CAS Key Laboratory of Microscale Magnetic Resonance, University of Science and Technology of China, Hefei 230026, China}
\affiliation{Synergetic Innovation Center of Quantum Information and Quantum Physics, University of Science and Technology of China, Hefei 230026, China}

\author{Jiangfeng Du}
\email{djf@ustc.edu.cn}
\affiliation{Hefei National Laboratory for Physical Sciences at the Microscale and Department of Modern Physics, University of Science and Technology of China, Hefei 230026, China}
\affiliation{CAS Key Laboratory of Microscale Magnetic Resonance, University of Science and Technology of China, Hefei 230026, China}
\affiliation{Synergetic Innovation Center of Quantum Information and Quantum Physics, University of Science and Technology of China, Hefei 230026, China}

\begin{abstract}
We theoretically propose an experimentally viable scheme to use an impurity atom in a dipolar Bose-Einstein condensate (BEC),
in order to probe analogue low-energy Lorentz violation from the modified dispersion at high energies 
as suggested by quantum theories of gravity.
We show that the density fluctuations in the dipolar BEC possess a Lorentz-violating Bogoliubov spectrum 
$\omega_\bk=c_0|\bk|f(c_0|\bk|/M_\star)$, with recovery of approximate Lorentz invariance (LI) at energy scales much below $M_\star$. 
When $f$ is adjusted to dip below unity somewhere, the impurity, analogously dipole coupled to the density fluctuations,
experiences analogue drastic Lorentz violation at arbitrarily low energies, reproducing the same responds of Unruh-DeWitt detector to 
Lorentz-violating quantum fields. Being a fundamentally quantum mechanical device, our quantum fluid platform provides an experimentally
realizable test field to verify whether the effective low energy theory can reveal unexpected imprints of the theory's high energy structure, 
in quantum field theory.

\end{abstract}

\baselineskip=0.45 cm
\maketitle
\newpage

\section{Introduction} 
Lorentz invariance (LI) is one of the fundamental symmetries of relativity, however, and may be challenged at 
sufficiently high energies suggested by many theories of quantum gravity \cite{AmelinoCamelia:2008qg}. 
As a phenomenal embodiment of quantum gravity,  any discovery of Lorentz violation thus would be 
an important signal of beyond standard model physics. Testing LI has recently given rise to 
widespread interest in diverse physical areas, spanning atomic physics, nuclear physics, high-energy physics, 
relativity, and astrophysics, see Ref. \cite{PhysRevD.55.6760, PhysRevD.58.116002, Mattingly:2005re, RevModPhys.83.11}, and references therein.

In the LI violation theory, there may be an explicit energy scale, such as the Planck energy $M_\ast$, which characterizes the violation and 
at energies much below which the LI is preserved approximately. The Planck energy, about $10^{19}~\text{GeV}$,
is much larger than any currently experimentally accessible energy scales, e.g., $10^{11}~\text{GeV}$ for the trans-\text{GZK} cosmic rays
that is the highest known energy of particles \cite{Mattingly:2005re}. In this regard,
direct observation of Planck scale Lorentz violation seems impossible in any experiments. 
Fortunately, strong Planck scale Lorentz violation could yield a small amount of violation at much lower energies,
and thus one can expect that the effective low-energy theory, for example in quantum field theory, 
can reveal unexpected imprints of the theory's high-energy 
structure \cite{Mattingly:2005re, PhysRevLett.93.191301, Polchinski_2012}.

Recently, it has been found that the transition behavior of 
low-energy Unruh-DeWitt detectors \cite{PhysRevD.14.870, Dewitt1979General, birrell_davies_1982, 10.1143/PTP.88.1, RevModPhys.80.787, Hu_2012} 
is ultra-sensitive to high-energy effects, such as polymer quantum field theories \cite{Kajuri_2016, KAJURI2018412, PhysRevLett.116.061301, PhysRevD.97.025008}. It has also been shown that the attainable sensitivity to nonlocal field theories 
is expected to outperform that of LHC experiments by many orders of magnitude \cite{PhysRevD.94.061902}.
Therefore, this feature of Unurh-DeWitt detectors may have highly potential application for 
the falsifiability of theoretical proposals, e.g., quantum gravity \cite{PhysRevLett.116.061301, PhysRevD.97.025008, PhysRevD.94.061902}. 
However, there is still no concrete experimental 
setup as a precursor to verify whether the Unurh-DeWitt detector indeed will work as expected before its future realistic application.

In this paper, we aim at closing this gap and propose to study the relevant physics 
with an experimentally accessible platform consisting of a dipolar BEC \cite{BARANOV200871} 
and an immersed impurity \cite{PhysRevLett.94.040404, PhysRevLett.91.240407}. 
From the perspective of analog, the density fluctuations in the dipolar BEC possessing a
roton spectrum \cite{PhysRevLett.98.030406, PhysRevA.97.063611, PhysRevLett.118.130404, PhysRevA.73.031602, PhysRevLett.90.250403},
due to the dipole-dipole interaction (DDI) between atoms (see below for a detailed discussion), 
are modeled as Lorentz-violating quantum fields. The impurity, 
analogously dipole coupled to the density fluctuations in the condensate, 
is modeled as an Unruh-DeWitt detector coupling to the Lorentz-violating quantum fields.
This specific quantum simulator in principle allows us to test a possible manifestation of the
effects caused by high energy structure in the low-energy quantum detection---drastic low-energy Lorentz violation in the impurity 
modeled as Unruh-DeWitt detector---as anticipated in Ref. \cite{PhysRevLett.116.061301}. 

Our paper is constructed as follows. In Sec. \ref{section2} we simply introduce our model---dipolar Bose-Einstein condensate, and
show how the Lorentz-violating quantum fields are simulated with the density fluctuation of the condensate. In Sec. \ref{section3}
we show how to simulate the Unruh-DeWitt detector with an impurity immersed in the dipolar BEC. In Sec. \ref{section4} we study 
the spontaneous excitation of the detector with inertial trajectory. Experimental feasibility of 
the relevant simulation within the current technologies of 
dipolar BEC is discussed in Sec. \ref{section5}. Finally, a summary of the main results of our work is present in Sec. \ref{section6}.

\section{Simulating Lorentz-violating quantum fields in dipolar Bose-Einstein condensate} \label{section2}
Let us begin with an interacting Bose gas comprising atoms or molecules of mass $m$, whose Lagrangian density 
is given by $(\hbar=1)$
\begin{eqnarray}\label{Lagrangian}
\nonumber
\mathcal{L}&=&\frac{i}{2}(\Psi^\ast\partial_t\Psi-\partial_t\Psi^\ast\Psi)-\frac{1}{2m}|\nabla\Psi|^2-V_\text{ext}|\Psi|^2
\\
&&-\frac{1}{2}|\Psi|^2\int\,d^3\bR^\prime\,V_\text{int}(\bR-\bR^\prime)|\Psi(\bR^\prime)|^2,
\end{eqnarray}
where $\bR=(\br, z)$ are spatial three-dimensional coordinates. The system is trapped by an external potential of the form 
$V_\text{ext}(\bR)=m\omega^2\br^2/2+m\omega^2_zz^2/2$. We will assume that over the whole time evolution
the gas is strongly confined in the $z$ direction, with aspect ratio $\kappa=\omega_z/\omega\gg1$. 
The two-body interaction contains two terms
\begin{eqnarray}\label{two-body-interaction}
V_\text{int}(\bR-\bR^\prime)=g_c\delta^3(\bR-\bR^\prime)+V_\text{dd}(\bR-\bR^\prime),
\end{eqnarray}
where $g_c$ is the contact interaction coupling, and 
\begin{eqnarray}
V_\text{dd}(\bR-\bR^\prime)=\frac{3g_d}{4\pi}\frac{[1-3(z-z^\prime)^2/|\bR-\bR^\prime|^2]}{|\bR-\bR^\prime|^3}
\end{eqnarray}
describes the dipolar interaction with coupling constant $g_d$.
Here the dipoles have been assumed to be polarized along the $z$ direction (or perpendicular to the $x$-$y$ plane) by an external field.
In addition, $g_c$ and $g_d$ can be controlled as required in the experiment. 
Let us note that it is the interaction 
between atoms or molecules (including both the contact and dipolar interaction) that results in 
the Lorentz-violating quasiparticle spectrum, similar to the case in Ref. \cite{EDWARDS2018319}, 
as shown in the following.
To guarantee the stability in the DDI-dominated regime \cite{PhysRevA.73.031602}, 
we impose that the system remains sufficiently close to 
the quasi-two-dimensional (quasi-2D) regime during the whole physical process considered.
We thus assume that in the $z$ direction, the condensate density has a Gaussian form,
$\rho_z(z)=(\pi\,d^2_z)^{-1/2}\exp[-z^2/d^2_z]$, with $d_z=\sqrt{1/m\omega_z}$. 
Integrating out the $z$ dependence, we can obtain 
the effective quasi-2D interaction, 
$V^\text{2D}_\text{int}(\br-\br^\prime)=\int\,dzdz^\prime\,V_\text{int}(\bR-\bR^\prime)\rho_z(z)\rho_z(z^\prime)$ \cite{PhysRevLett.118.130404}.

In the 2D case, we decompose the 2D field operator as 
\begin{eqnarray}
\hat{\psi}=\psi_0(1+\hat{\phi}) 
\end{eqnarray}
with $\psi_0=\sqrt{\rho_0}e^{i\theta_0}$, where $|\psi_0(\br)|^2=\rho_0\simeq\mathrm{const}$ represents 
the 2D condensate density, and $\hat{\phi}$ describes the perturbations (excitations) on the top of the condensate. 
The Bogoliubov-de Gennes equation 
for the fluctuation field $\hat{\phi}$ reads \cite{2001camw.book1C, PhysRevA.97.063611}
\begin{eqnarray}\label{Bogoliubov-eq}
\nonumber
i\partial_t\hat{\phi}&=&-\frac{1}{2m}\nabla^2_\br\hat{\phi}
+\rho_0\int\,d^2\br^\prime\,V^\text{2D}_\text{int, 0}(\br-\br^\prime)
\\
&&\times\big[\hat{\phi}(\br^\prime)+\hat{\phi}^\dagger(\br^\prime)\big],
\end{eqnarray}
where the condensate has been assumed to be 
static, i.e., $\mathbf{v}=\frac{1}{m}\nabla_\br\theta_0=0$.
By solving Eq. \eqref{Bogoliubov-eq}, we can write the density fluctuations in Heisenberg representation as 
\begin{eqnarray}\label{DF}
\nonumber
\delta\hat{\rho}(t,\br)&\simeq&\rho_0(\hat{\phi}+\hat{\phi}^\dagger)
\\  \nonumber
&=&\sqrt{\rho_0}\int[d\bk/(2\pi)^2](u_\bk+v_\bk)\times[\hat{b}_\bk(t)e^{i\bk\cdot\br}
\\
&&+\hat{b}^\dagger_\bk(t)e^{-i\bk\cdot\br}], 
\end{eqnarray}
where the Bogoliubov quasiparticle operators 
$\hat{b}_\bk(t)=\hat{b}_\bk\,e^{-i\omega_\bk\,t}$ satisfy the usual Bose commutation rules 
$[\hat{b}_\bk, \hat{b}_{\bk^\prime}^\dagger]=(2\pi)^2\delta^2(\bk-\bk^\prime)$. 
Bogoliubov parameters are given by
\begin{eqnarray}
\nonumber
u_\bk&=&(\sqrt{\cH_\bk}+\sqrt{\cH_\bk+2\cA_\bk})/2(\cH^2_\bk+2\cH_\bk\cA_\bk)^{1/4},
\\
v_\bk&=&(\sqrt{\cH_\bk}-\sqrt{\cH_\bk+2\cA_\bk})/2(\cH^2_\bk+2\cH_\bk\cA_\bk)^{1/4},
\end{eqnarray}
 and the quasiparticle frequency 
$\omega_\bk=\sqrt{\cH^2_\bk+2\cH_\bk\cA_\bk}$ with 
$\cH_\bk=k^2/2m$ and $\cA_\bk=\rho_0V^\text{2D}_\text{int, 0}(k)$ \cite{PhysRevA.97.063611}. 
Here $k=|\bk|$ and the Fourier transformation of the effective quasi-2D interaction is given by \cite{PhysRevA.73.031602}
\begin{eqnarray}
V^\text{2D}_\text{int, 0}(k)=g^\text{eff}_0(1-\frac{3R}{2}kd_zw[\frac{kd_z}{\sqrt{2}}]), 
\end{eqnarray}
with $w[x]=\exp[x^2](1-\erf[x])$,
an effective contact coupling $g^\text{eff}_0=\frac{1}{\sqrt{2\pi}d_z}(g_c+2g_d)$, and the dimensionless ratio is defined as
\begin{eqnarray}
R=\sqrt{\pi/2}/(1+g_c/2g_d).
\end{eqnarray}
Note that the parameter $R$ could be tunable via Feshbach resonance \cite{PhysRevLett.81.69, Inouye1998Observation} and rotating polarizing field \cite{PhysRevLett.89.130401}. It ranges from $R=0$ (when $g_d/g_c\rightarrow0$, i.e., contact dominance), to $R=\sqrt{\pi/2}$ (when $g_d/g_c\rightarrow\infty$, i.e., DDI dominance).

\begin{figure}
\centering
\includegraphics[width=0.3\textwidth]{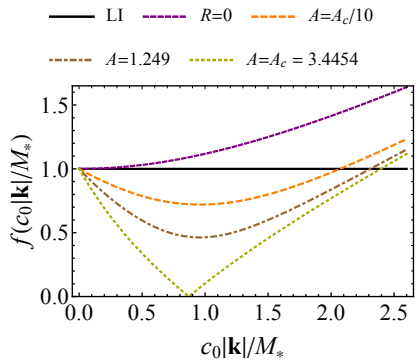}
\caption{(Color online) The dimensionless function $f$ shown in \eqref{Dispersion} as a function of $c_0|\bk|/M_\ast$.
LI is the Lorentz invariant case where $f=1$. $R=0$ denotes the contact interaction case, where $f$ is independent of $A$. 
For DDI dominance, $R=\sqrt{\pi/2}$, $f$ dips below 1 for an interval of $c_0|\bk|/M_\ast$. Note that 
$f$ becomes negative when $A>A_c=3.4454$, which means the spectrum of quasiparticle 
becomes unstable.}\label{fig1}
\end{figure}

The density fluctuations shown in Eq. \eqref{DF} closely resemble a Lorentz-violating scalar field with an
explicit dispersion relation given by 
\begin{eqnarray}\label{Dispersion}
\nonumber
\omega_\bk&=&c_0|\bk|\sqrt{1-\frac{3R}{2}\sqrt{A}\frac{c_0|\bk|}{M_\ast}w\bigg[\sqrt{\frac{A}{2}}\frac{c_0|\bk|}{{M_\ast}}\bigg]
+\frac{1}{4}\frac{c^2_0|\bk|^2}{M^2_\ast}}
\\
&=&c_0|\bk|f(c_0|\bk|/M_\ast),
\end{eqnarray}
where $c_0=\sqrt{g^\text{eff}_0\rho_0/m}$ is the speed of sound, another dimensionless parameter 
\begin{eqnarray}
A=g^\text{eff}_0\rho_0/\omega_z
\end{eqnarray}
represents the effective chemical potential as measured relative to the transverse trapping,  and 
\begin{eqnarray}
M_\ast=mc^2_0
\end{eqnarray}
is the analog energy scale of Lorentz violation. In the 
DDI dominance regime $A$ should be assumed to be not larger than the critical value $A_c=3.4454$, since beyond which 
the spectrum of quasiparticle becomes unstable \cite{PhysRevA.97.063611, PhysRevA.73.031602, PhysRevLett.118.130404}.
Therefore, to keep the stability of the spectrum of quasiparticle in the 
DDI dominance regime, $A\leq\,A_c$ will be taken throughout the whole paper and this condition 
requires that the frequency of the confinement along the $z$-direction satisfies 
$\omega_z\geq\frac{2mg^2_d\rho^2_0}{\pi\,A^2_c}$. The dispersion relation \eqref{Dispersion} is approximately 
Lorentz invariant $(f(c_0|\bk|/M_\ast)\simeq1)$ for $c_0|\bk|/M_\ast\ll1$. 
By appropriately setting the relevant parameters $A$ and $R$, 
the dispersion could be analogously superluminal $(f(c_0|\bk|/M_\ast)>1)$ and subluminal 
$(f(c_0|\bk|/M_\ast)<1)$. In Fig. \ref{fig1}, we plot the function $f(c_0|\bk|/M_\ast)$ shown in \eqref{Dispersion}
to see how the LI is violated in the dispersion.
For the DDI dominance, $R=\sqrt{\pi/2}$, the analogous subluminal spectrum develops a roton minimum 
for sufficiently large $A$, and the LI is strongly broken near $c_0|\bk|/M_\ast\simeq0.9$ \cite{PhysRevA.97.063611}. Alternatively, 
$f(c_0|\bk|/M_\ast)$ could dip below $1$ for an interval of $c_0|\bk|/M_\ast$.
This feature will be very helpful for us to investigate in quantum field theory, 
how to reveal unexpected imprints of the theory's high energy structure with well-tested low energy detection in the following.

\section{Unruh-DeWitt detector model in dipolar BEC} \label{section3}
\begin{figure}
\centering
\includegraphics[width=0.46\textwidth]{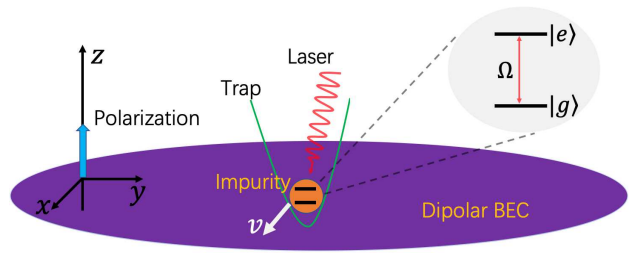}
\caption{(Color online) Schematic of the moving (with speed parameter $v$) impurity with effective 
internal frequency $\Omega$ immersed in a quasi-2D dipolar condensate (in purple).
Here the dipoles of atoms or molecules have been assumed to be polarized along the $z$ direction or perpendicular to the $x$-$y$ plane (in blue).}\label{detector}
\end{figure}

In quantum field theory,
usually the field is probed with a linearly coupled two-level (1 and 2) Unruh-DeWitt detector 
\cite{PhysRevD.14.870, Dewitt1979General, birrell_davies_1982, 10.1143/PTP.88.1, RevModPhys.80.787, Hu_2012}.
This detector model captures the essential features of the light-matter interaction 
when angular momentum interchange is negligible \cite{PhysRevD.87.064038, PhysRevA.89.033835}.
Inspired by the seminal atomic quantum dot idea introduced in Refs. \cite{PhysRevLett.94.040404, PhysRevLett.91.240407}, 
we model an impurity consisting of a two-level ($1$ and $2$) atom as the Unruh-DeWitt detector, and assume the impurity is immersed in 
the quasi-2D dipolar BEC discussed above (see Fig. \ref{detector}). The impurity's motion is assumed to 
be externally imposed by a tightly confining and relatively moving trap potential,
so that we can focus only on its internal degrees of freedom.

In this detector model, a monochromatic external electromagnetic field at frequency $\omega_L$, which is close 
to resonance with the $1\rightarrow2$ transition $\omega_L\simeq\omega_{21}$, is applied to illuminate 
the impurity, and correspondingly the Rabi frequency is $\Omega$.
We can write the system Hamiltonian in the form
\begin{eqnarray}\label{Hamiltonian}
\nonumber
H(t)&=&H_C+H_A(t)=\sum_{\bf k\neq0}\omega_{\bf k}\hat{b}^\dagger_{\bf k}\hat{b}_{\bf k}
\\    \nonumber
&&+\omega_{21}|2\rangle\langle2|-\bigg(\frac{\Omega}{2}e^{-i\omega_Lt}|2\rangle\langle1|+\mathrm{H.c.}\bigg)
\\
&&+\sum_sg_s\hat{\rho}(\br_A(t))|s\rangle\langle\,s|,
\end{eqnarray}
where the last term is the collisional coupling between the impurity and Bose gas.
$\hat{\rho}(\br_A)=\hat{\psi}^\dagger(\br_A)\hat{\psi}(\br_A)$ denotes the field density operator of the atomic Bose gas, and 
$\br_A(t)$ is the time-dependent impurity position. $g_s$ are the interaction constant 
between the impurity in $s=1, 2$ state and the condensate. In the rotating frame,
the detector's Hamiltonian including its interaction with the Bose gas can be rewritten as 
\begin{eqnarray} \label{detector-interaction}
\nonumber 
H_A(t)&=&\omega_{21}|2\rangle\langle2|-\frac{1}{2}\omega_L(|2\rangle\langle2|-|1\rangle\langle1|)
-\frac{1}{2}\Omega(|2\rangle\langle1|
\\
&&+\mathrm{H.c.})+\sum_sg_s\hat{\rho}(\br_A(t))|s\rangle\langle\,s|.
\end{eqnarray}
Then, using the rotated $|g, e\rangle=(1/\sqrt{2})(|1\rangle\pm|2\rangle)$ basis and defining $g_\pm=\frac{1}{2}(g_1\pm\,g_2)$, 
we can further rewrite the Hamiltonian \eqref{detector-interaction} as
\begin{eqnarray} \label{detector-interaction2}
H_A(t)&=\frac{\Omega}{2}\sigma_z+\frac{\delta}{2}\sigma_x+\hat{\rho}(\br_A(t))[g_++g_-\sigma_x],
\end{eqnarray}
where $\sigma_z$ and $\sigma_x$ are the conventional Pauli matrices, and $\delta=\omega_L-\omega_{21}$ is the detuning. 
In this rotated basis, the Rabi frequency $\Omega$ determines the splitting between the $|g, e\rangle$ states, while the detuning 
$\delta$ gives a coupling term.

Let us note the last term of the above Hamiltonian \eqref{detector-interaction2}
denotes the interaction between the impurity and Bose gas.
This interaction contains two terms: the first one proportional to $g_+$ is similar to the reminiscent 
coupling of a charged particle to an electric field, 
while the other resembles a standard electric-dipole coupling mediated by a coupling constant $g_-$. 
By suitably choosing the internal atomic states and properly tuning the interaction constants 
(e.g., via Feshbach resonances \cite{PhysRevLett.81.69, Inouye1998Observation}), 
the first term could be cancelled as a result of $g_+=0$ \cite{PhysRevA.74.041605, PhysRevA.77.052705, RevModPhys.82.1225}, behaving like
the analog charge neutrality. Furthermore, the atomic density operator $\hat{\rho}(\br)$, as shown above, can be split into 
its average value $\rho_0$ and small fluctuations $\delta\hat{\rho}(\br)$ in \eqref{DF}. With the suitable 
detuning $\delta$ between driving frequency and the impurity's internal level space, one can 
exactly compensate the coupling to the average density, $\delta/2+g_-\rho_0=0$ \cite{PhysRevLett.118.045301, marino2020zeropoint}.
Under all these assumptions, the impurity's Hamiltonian including the its interaction with the condensate 
can finally be written as 
\begin{eqnarray} \label{detector-interaction3}
H_A(t)&=\frac{\Omega}{2}\sigma_z+g_-\sigma_x\delta\hat{\rho}(\br_A).
\end{eqnarray}
The coupling of the impurity to the condensate in the Hamiltonian \eqref{detector-interaction3}
 has the analogous form $g_-\sigma_x\delta\hat{\rho}(\br_A)$ 
of a two-level atom dipole coupled to the quantum Lorentz-violating scalar 
field at its position $\br_A$.

\section{Spontaneous excitation of inertial detector}\label{section4}

\begin{figure*}
\centering
\subfigure[~~\normalsize{$\Omega/M_\ast=0.3$}]{\includegraphics[width=0.28\textwidth]{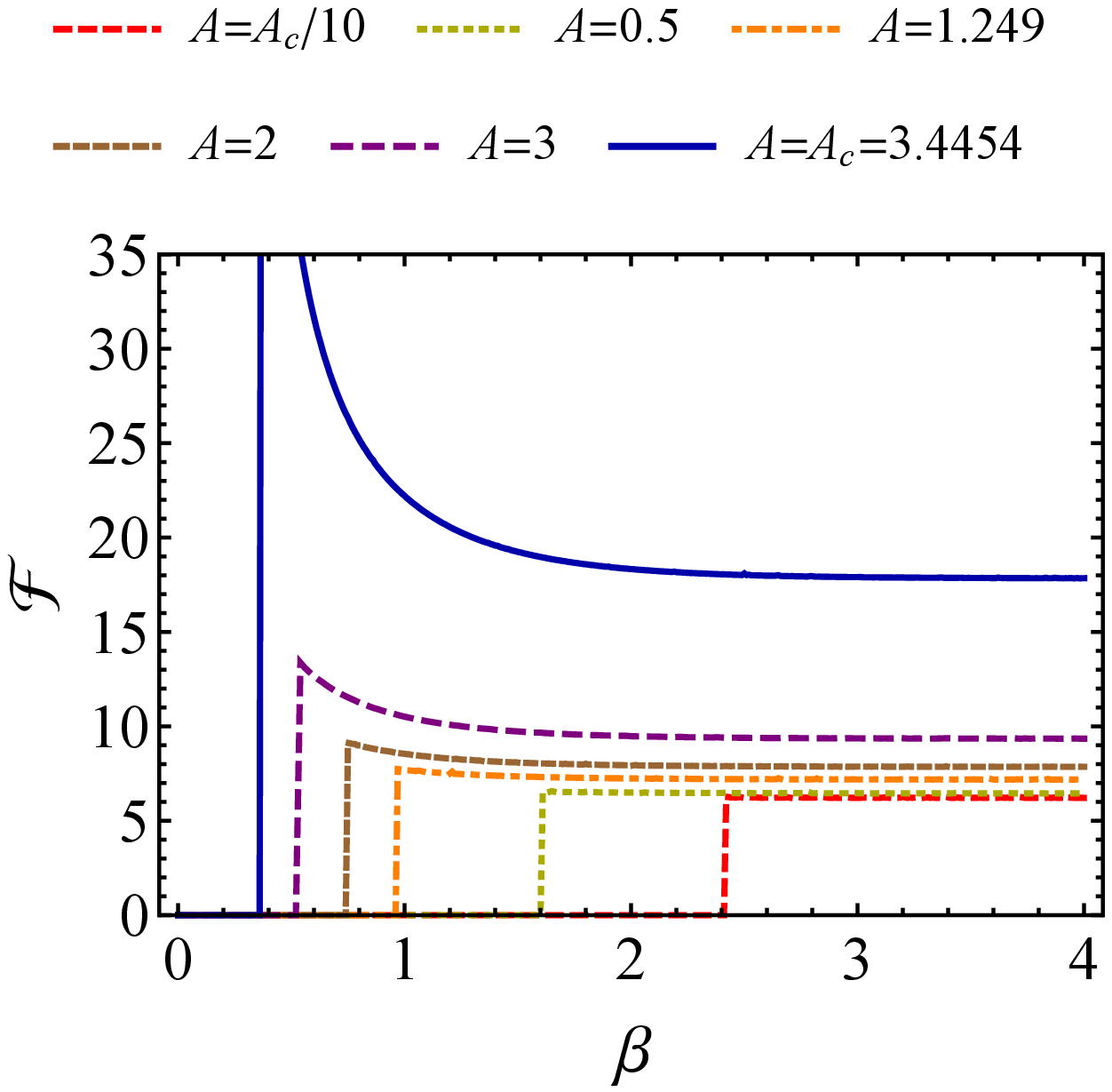}}
\hspace*{0.5em}
\subfigure[~~\normalsize{$\Omega/M_\ast=0.1$}]{\includegraphics[width=0.28\textwidth]{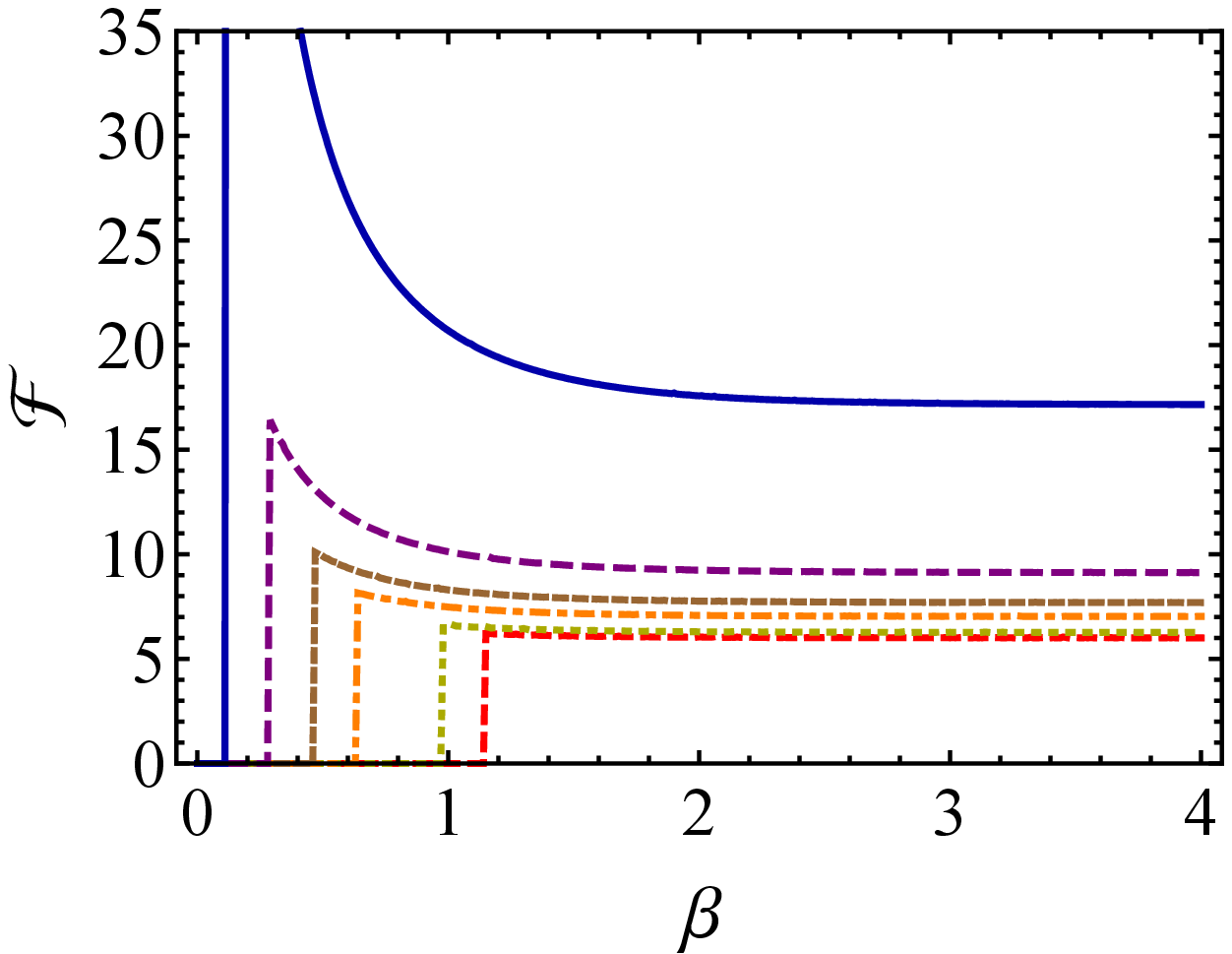}}
\hspace*{0.5em}
\subfigure[~~\normalsize{$\Omega/M_\ast=0.01$}]{\includegraphics[width=0.28\textwidth]{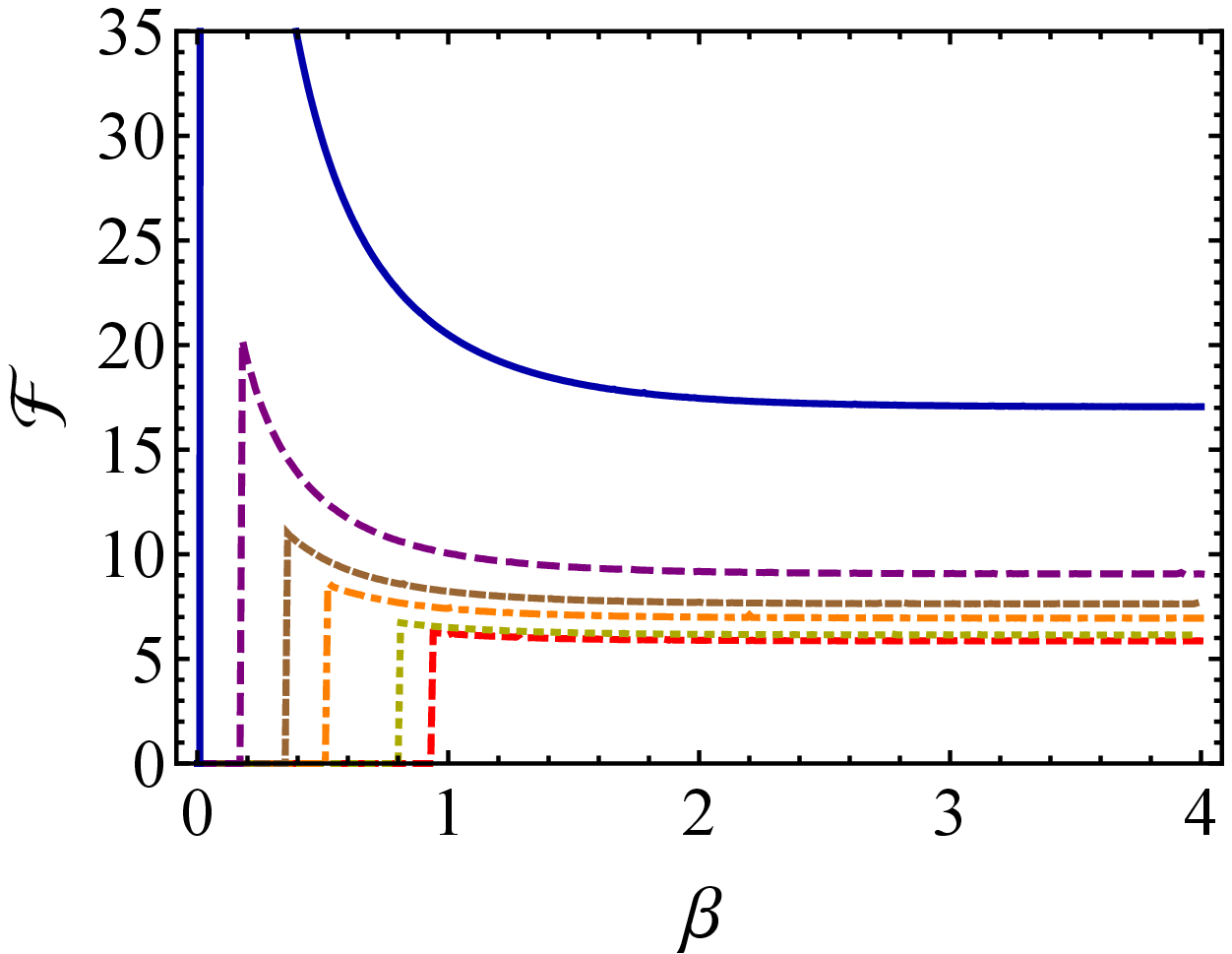}}
\caption{(Color online) Transition rate shown in \eqref{TR} for excitations $(\Omega/M_\ast>0)$ vs $\beta$ for different effective chemical potential $A$, in units of $\frac{\rho_0M_\ast\,g^2_-}{(2\pi)^3c^{-2}_0}$. Here we assume that the DDI dominance case ($R=\sqrt{\pi/2}$) 
is valid.}\label{fig2}
\end{figure*}

Consider the above impurity detector with the Hamiltonian given in \eqref{detector-interaction3},
we assume that it is pointlike and moves on the worldline $(t(\tau), {\bf r}(\tau))$, where $\tau$ is the proper time.
In the detector's frame, the coupling between the impurity and the density fluctuations of the condensate (or Bogoliubov field)
in the interaction picture can be rewritten as 
\begin{eqnarray}
H_{int}(\tau)=g_-\big(e^{i\Omega\tau}\sigma_++e^{-i\Omega\tau}\sigma_-\big)\delta\hat{\rho}(t(\tau),\br_A(\tau)).
\end{eqnarray}
We take the initial state of joint system (the impurity and the density fluctuations) 
before the interaction to be $|0\rangle\otimes\,|g\rangle$. We pay attention to 
the probability for the detector to be excited, i.e., at state $|e\rangle$, as a result of the interaction.
Since here the coupling constant $g_-$ is treated as a small parameter, it enables this probability to be calculated 
to first order in perturbation theory in $g_-$ \cite{birrell_davies_1982, 10.1143/PTP.88.1, RevModPhys.80.787, Hu_2012, Louko_2008, Satz_2007, Schlicht_2004}, 
\begin{eqnarray} 
\nonumber
P(\Omega)&=&\big|\langle\Psi\,e|i g_-\int^\infty_{-\infty}d\tau\big(e^{i\Omega\tau}\sigma_++e^{-i\Omega\tau}\sigma_-\big)
\\  \nonumber
&&\otimes\delta\hat{\rho}(t(\tau),\br_A(\tau))|g\,0\rangle\big|^2
\\ 
&=&g_-^2\int^\infty_{-\infty}d\tau^\prime\int^\infty_{-\infty}d\tau^{\prime\prime}e^{-i\Omega(\tau^\prime-\tau^{\prime\prime})}
W(\tau^\prime,\tau^{\prime\prime}),
\end{eqnarray}
where $|\Psi\rangle$ is quantum state of the Bogoliubov field satisfying $|\Psi\rangle\langle\Psi|+|0\rangle\langle0|={\bf 1}$, and 
$W(\tau^\prime,\tau^{\prime\prime})=\langle0|\delta\hat{\rho}(t(\tau^\prime),\br_A(\tau^\prime))\delta\hat{\rho}(t(\tau^{\prime\prime}),\br_A(\tau^{\prime\prime}))|0\rangle$ denotes the Wightman function of the Bogoliubov field. We can consider situations in which
the Wightman function is stationary, i.e., $W(\tau^\prime,\tau^{\prime\prime})=W(\tau^\prime-\tau^{\prime\prime})$. 
It implies that the correlations between two events depend only on the difference of times between them. Then 
we may convert $P(\Omega)$ into the transition rate per unit time, as done in Refs. \cite{ Louko_2008, Satz_2007, Schlicht_2004}. 
After some straight transformations and calculations \cite{Louko_2008, Satz_2007, Schlicht_2004}, 
the transition rate per unit time finally is given by 
\begin{eqnarray} \label{T-rate}
\mathcal{F}(\Omega)=g^2_-\int^\infty_{-\infty}dse^{-i\Omega\,s}W(s, 0),
\end{eqnarray}
where $s=\tau^\prime-\tau^{\prime\prime}$. We will use Eq. \eqref{T-rate} in the following study.

We consider an inertial detector with the trajectory given by $(t(\tau), \br(\tau))=(\tau, c_0\tau\tanh\beta, 0)$,
where $c_0\tanh\beta$ denotes the velocity with respect to the distinguished inertial frame, and for simplicity is set to be positive. 
Note that the detector's velocity has been assumed to be smaller than the sound speed, $c_0$.
Assuming both the impurity and the Bogoliubov field 
initially in the ground state and working in the frame comoving with the impurity, we obtain 
the detector's transition rate by substituting Eq. \eqref{DF} into Eq. \eqref{T-rate},
\begin{eqnarray}\label{TR}
\nonumber
\mathcal{F}(\Omega)&=&\int^\infty_{-\infty}ds\int\,d\bk\frac{\rho_0g^2_-}{(2\pi)^4}\frac{\cH_\bk}{\omega_\bk}
e^{-i(\Omega+\omega_\bk-c_0k_x\tanh\beta)s}
\\     \nonumber
&=&\frac{\rho_0M_\ast\,g^2_-}{(2\pi)^3c^{-2}_0}\int^\infty_0dg\frac{g^2}{f(g)}
\\   
&&\times\frac{\Theta\big(g\tanh\beta-|\Omega/M_\ast+gf(g)|\big)}{\sqrt{g^2\tanh^2\beta-\big(\Omega/M_\ast+gf(g)\big)^2}},
\end{eqnarray}
where $g=c_0|\bk|/M_\ast$, $\Omega$ denotes the effective energy space 
between the impurity's two levels, and $\Theta$ is the Heaviside function. In the low-speed limit, i.e., when $\beta\ll1$, we have $\tanh\beta\thickapprox\beta$, 
then the effective-energy expression in Eq. \eqref{TR} $\Omega+\omega_{\bf k}-c_0k_x\tanh\beta$ is approximate to 
$\Omega+\omega_{\bf k}-c_0k_x\beta$, and the detector's transition rate could be given by
\begin{eqnarray}
\nonumber 
\mathcal{F}(\Omega)
&=&\frac{\rho_0M_\ast\,g^2_-}{(2\pi)^3c^{-2}_0}\int^\infty_0dg\frac{g^2}{f(g)}
\\
&&\times\frac{\Theta\big(g\beta-|\Omega/M_\ast+gf(g)|\big)}{\sqrt{g^2\beta^2-\big(\Omega/M_\ast+gf(g)\big)^2}}.
\end{eqnarray}

In Eq. \eqref{TR} the crucial issue, resulting from the definition of Heaviside function, is the behavior of the argument of $\Theta$:
Under what condition is the argument of $\Theta$ positive for at least some interval of $g$? 
If $f(g)\geq1$ for all $g$, the argument of $\Theta$ clearly is always negative for all positive $\Omega$,
and thus the corresponding transition rate 
$\mathcal{F}(\Omega)$ vanishes. The vanishing transition rate suggests the detector does not become spontaneously excited.
Note that $f(g)=1$ corresponds to the usual massless scalar field (Lorentz-invaiant field) case,
thus the Lorentz-invariant vacuum field induces no spontaneous excitations in the inertial detector \cite{birrell_davies_1982}.

An interesting case is when $f(g)$ dips somewhere below unity \cite{PhysRevLett.116.061301, PhysRevD.97.025008}. 
Concretely, we assume $f_c=\inf\, f$, and $0<f_c<1$ (see more details in Fig. \ref{fig1}). 
In this scenario, the detector behaves quite differently for 
rapidities below and above the critical value $\beta_c=\arctanh(f_c)$. 
When $0<\beta<\beta_c$, we see from the argument of 
$\Theta$ in \eqref{TR} that $\mathcal{F}(\Omega)$ vanishes for $\Omega>0$. 
It implies that the detector remains unexcited. 
This result is consistent with that for the uniformly moving detector coupled to a massless scalar field \cite{birrell_davies_1982}.
However, when $\beta>\beta_c$, the argument of $\Theta$ in \eqref{TR} keeps 
positive for $0<\Omega/M_\ast<\sup_{g\geq0}g[\tanh\beta-f(g)]$, and thus $\mathcal{F}(\Omega)$ does not vanish.
It suggests that the detector gets spontaneously excited, at arbitrarily small positive $\Omega$. 
This result is quite different from that of the uniformly moving detector coupled to the massless scalar field, 
which never gets spontaneously excited \cite{birrell_davies_1982}. Besides, 
when $\Omega<0$, no matter the rapidities are below or above the critical 
value, the detector has a nonvanishing deexcitation rate that depends on the rapidities.

We give a brief summary here: The Bogoliubov spectrum in \eqref{Dispersion} violates Lorentz invariance,
but is approximately Lorentz invariant as $c_0|\bk|/M_\ast$ approaches to zero.
Furthermore, its smooth positive-valued function $f$ could dip somewhere below unity,
satisfying $0<f_c<1$ with $f_c=\inf\, f$. Then, the broken LI, strongly at the energy scale $M_\ast$, may induce 
the inertial Unruh Dewitt detector with rapidity $\beta>\beta_c=\arctanh(f_c)$
in the preferred frame to experience spontaneous excitations and 
deexcitations at arbitrarily low $|\Omega|$. The anticipation in Ref. \cite{PhysRevLett.116.061301} could be verified in 
the analogue gravity system proposed here.

\begin{figure}
\centering
\includegraphics[width=0.3\textwidth]{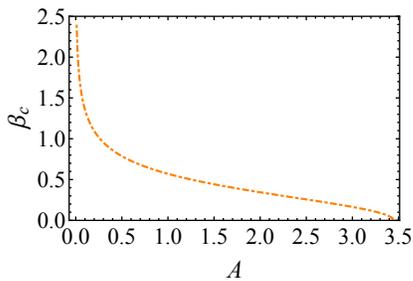}
\caption{(Color online) The critical rapidity $\beta_c$ as a function of the effective chemical potential $A$ for the DDI dominance case ($R=\sqrt{\pi/2}$).}\label{fig3}
\end{figure}

In Fig. \ref{fig2}, we take the dispersion relation \eqref{Dispersion} in the DDI domination regime ($R=\sqrt{\pi/2}$)
and plot the transition rate $\mathcal{F}(\Omega)$ in \eqref{TR}, as a function of $\beta$.
Clearly, the transition rate remains vanishing in the beginning, then suddenly becomes a maximal value when the rapidity $\beta$ 
exceeds the critical $\beta_c$, and decays with the increase of the rapidity. 
Moreover, in the DDI domination regime how the critical rapidity $\beta_c$ 
changes with the effective chemical potential $A$ is shown in Fig. \ref{fig3}.
Since $f_c$ in our setup could approach to zero when working in the DDI dominance regime for
sufficiently large $A$ (see Fig. \ref{fig1}), correspondingly, the critical rapidity 
$\beta_c=\arctanh(f_c)$ beyond which the detector gets spontaneously excited approaches to zero as well.
It suggests that even for a quite small rapidity, one can still observe the spontaneous excitations of inertial detector 
as a result of the broken LI of quantum fields. This ultralow-speed demand could reduce the experimental challenge.

\section{Experimental implementation}\label{section5}

So far, the experimental realization of magnetic DDI-dominated condensate \cite{BARANOV200871} has been reported with different 
atoms \cite{Chromium, PhysRevLett.108.210401, PhysRevLett.107.190401}. Besides, the realization of BEC made up of molecules 
with permanent electric dipoles \cite{doi:10.1021/cr300092g} is now at the forefront of ongoing research; 
cf., e.g., Refs. \cite{PhysRevLett.114.205302, PhysRevLett.116.205303, PhysRevLett.119.143001, De-Marco853, Son}.
As a result of the DDI between atoms, the excitation spectrum of dipolar BEC displays a deviation from
the Lorentz invariant one, and even displays a roton minimum when the BEC is in the DDI-dominated regime \cite{PhysRevA.97.063611, PhysRevLett.118.130404, PhysRevA.73.031602, PhysRevLett.90.250403, PhysRevLett.98.030406}. 
Experimental observation of roton modes in ultracold dipolar quantum gases has been reported recently 
\cite{Chomaz, DDIexperiment, PhysRevLett.123.050402, PhysRevLett.122.183401}.
The density fluctuations in the dipolar BEC, as shown above, thus could furnish a quantum simulation 
of the quantum field with broken LI, within current experimental reach.
In addition, great advances in high-precision measurements of correlation functions 
\cite{PMID:23907531, PMID:21350171, thermal-Hawking-radiation1, thermal-Hawking-radiation2}  
pave the way to explore the possible extraordinary propagation of quantum field due to the broken LI.

The hybrid systems \cite{RevModPhys.91.035001}, consisting of trapped ions and ultracold atomic systems, has recently 
emerged as a new platform for fundamental research in quantum physics. In experiment, a trapped single ion \cite{2010Natur.464..388Z} 
and single electron \cite{2013Natur.502..664B} coupled to a BEC have been implemented. Moreover, the impurities' dynamics can be used to 
probe the density profile of BEC \cite{PhysRevLett.105.133202, PhysRevLett.111.070401, PhysRevLett.109.235301, PhysRevLett.121.130403}.
By designing the external trap potential which tightly confines the impurity and moves uniformly, 
one can impose the comoving motion on the impurity. 
A quasiparticle formed by a mobile impurity interacting with a surrounding BEC, 
was observed in two parallel experiments \cite{PhysRevLett.117.055301, PhysRevLett.117.055302}, using different physical systems and techniques.
Thus, it is possible in principle, to design a hybrid system combining a impurity and a dipolar BEC,
in order to explore the dynamics of the impurity with the roton modes due to DDI between atoms. 
This proposed cold atom setup could be as the quantum simulator of the broken LI physics, as discussed above.

\section{Conclusions and discussions}\label{section6}

In this paper, we have proposed a cold atom setup where an external potential trapped impurity 
immersed in a dipolar BEC and coupled to its density fluctuations serves as a quantum simulator 
of a two-level atom dipole coupled to the quantum scalar field with broken LI. We
have shown that the Bogoliubov spectrum in the dipolar BEC deviates from the Lorentz invariant one,
and the deviation could be tunable via Feshbach resonance and rotating polarizing field.
Due to the broken LI, the inertial impurity modeled as a Unruh-DeWitt detector 
can be spontaneously excited when its rapidity exceeds the criticality $\beta_c=\arctanh(f_c)$.
Interestingly, the critical rapidity in our scenario can be controlled and even can approach to 
zero if the ratio between the atomic contact interaction and DDI is appropriate. 
Therefore, our proposal can experimentally demonstrate that the effective low energy theory can reveal unexpected 
imprints of the theory's high energy structure, in quantum field theory.

Note that the detector initially prepared at its ground state
could get excited when its velocity exceeds the critical one as discussed above, and 
the critical velocity is smaller than the speed of sound as a consequence of the spectrum function $f(c_0|\bk|/M_\ast)<1$. 
During the excitation process the moving detector could also simultaneously
create the excitations (or quasiparticles) in the dipolar BEC, which have the spectrum as shown 
in Eq. \eqref{Dispersion}. Therefore, one can also indirectly observe the created 
quasiparticles in the dipolar BEC by checking if the detector becomes excited, or by measuring the energy of the detector 
with the similar experimental methods in Refs. \cite{PhysRevLett.117.055301, PhysRevLett.117.055302}.

Our proposed quantum fluid platform has potential as a quantum simulator \cite{RevModPhys.86.153}
of quantum field theories: the tunability of the impurity's motion, of the sound speed, 
of the Bogoliubov spectrum form, of the condensate's dynamics and geometry, allows us in the experimentally accessible regime,
to explore open questions concerning Unruh effect \cite{PhysRevLett.101.110402}, and 
why its robustness to high energy modifications of the dispersion relation \cite{PhysRevLett.123.041601}
behaves differently from that of 
its equivalence principle dual---Hawking effect \cite{PhysRevD.51.2827}.
In addition, the impurity could also be used as a detector to explore   
the intricate many-body correlations between analogue high-energy quasiparticles in the dipolar BEC \cite{PhysRevA.97.063611} 
due to the DDI between atoms.

~~~~~
\begin{acknowledgments}

This work was supported by the National Key R\&D Program of China (Grant No. 2018YFA0306600), the CAS (Grants No. GJJSTD20170001 and No. QYZDY-SSW-SLH004), and Anhui Initiative in Quantum Information Technologies (Grant No. AHY050000).  
ZT was supported by the National Natural Science Foundation of 
China under Grant No. 11905218, and the CAS Key Laboratory for Research in Galaxies and Cosmology, Chinese Academy of Science (No. 18010203). 
\end{acknowledgments}


\bibliography{LIV}

\end{document}